# Using optical resonances to control heat generation and propagation in silicon nanostructures

*Stefano Danesi,[a,b] Ivano Alessandri[a,c,d*]*

[a] INSTM-UdR Brescia, via Branze 38, 25123 Brescia, Italy

[b] Department of Mechanical and Industrial Engineering, via Branze 38, 25123 Brescia, Italy

[c] Department of Information Engineering, University of Brescia, via Branze 38, 25123 Brescia, Italy

[d] INO-CNR, via Branze 38, 25123 Brescia, Italy

# ABSTRACT

Here we propose a new computational approach to light-matter interactions in silicon nanopillars, which simulates heat generation and propagation dynamics occurring in continuous wave laser processing over a wide temporal range (from 1 fs to about 25 hours). We demonstrate that a rational design of the nanostructure aspect ratio, type of substrate, laser irradiation time and wavelength enables amorphous-to-crystalline transformations to take place with a precise, sub-wavelength spatial localization. In particular, we show that visible light can be exploited to selectively crystallize internal region of the pillars, which is not possible by conventional treatments. A detailed study on lattice crystallization and reconstruction dynamics reveals that local heating drives the formation of secondary antennas embedded into the pillars, highlighting the importance of taking into account the spatial and temporal evolution of the optical properties of the material under irradiation. This approach can be easily extended to many types of nanostructured materials and interfaces, offering a unique computational tool for many applications involving opto-thermal processes (fabrication, data storage, sensing, catalysis, resonant laser printing, opto-thermal therapy etc…).

Three-dimensional integrated circuits (3D-ICs) offer promising routes for pushing the development of electronics beyond the limits of the Moore's law. (1-3) In this context, photonic devices based on 3D nanostructures can play a key role in fostering the technology breakthrough. (4-6) The fabrication of integrated photonic devices passes through an accurate control of silicon doping and lattice reconstruction, which are normally achieved by specific treatments involving multiple steps. For example, dopants are usually introduced into the silicon lattice by ion implantation. (7) This procedure yields local lattice damages and formation of point defects, which can further migrate, giving rise to uncontrolled segregations and extended dislocations. In many cases, silicon-based structures are also preliminary amorphized in order to prevent ion channeling. In both the cases, post-implantation damage recovery and silicon lattice reconstruction are realized by thermal treatment. However, conventional thermal treatment like furnace annealing and solid-phase epitaxial regrowth (SPER) are carried out at high temperature (T>600°C), which poses severe limitations for process design, as real devices include different circuit components that can undergo thermal degradation. (8,9) Laser-based thermal processing is a valuable tool to localize heating at specific positions, as the thermal effects can be limited to the range of few micrometers. In general, the optical annealing of silicon chips is realized by means of UV lasers ($\lambda$=266 nm). In the UV range silicon exhibits high ohmic losses, allowing the laser power to be uniformly distributed all over the chip. Because of its non-resonant nature, UV laser processing does not allow heat generation and propagation in silicon nanostructures to be controlled, which would require the extension of spatial resolution at the level of nanoscale. (10) On the other hand, at optical frequencies silicon is characterized by low ohmic losses, which makes heat generation ineffective. (11) However, recent works demonstrate that very strong local heating can be generated when the frequency of the laser beam is coupled with that of the optical resonance of a silicon nanostructure. Under these conditions, the local temperature can exceed 600°C, which is a typical temperature for inducing lattice reconstruction. (12) Moreover, the overall opto-thermal effects depend not only on optical excitation but also on thermal dissipation, which is deeply influenced by the nanostructure morphology. These effects have been mainly applied for Raman spectroscopy (13,14) and optical thermometry purposes, (15) but their direct exploitation for controlling silicon lattice reconstruction has not been devised so far.

Here we developed a theoretical and computational model which is able to predict the spatial and temporal evolution of lattice crystallization and reconstruction in silicon nanopillars irradiated with a CW visible laser ($\lambda$=532) and we compared these results with those obtained by conventional annealing processes performed by either furnaces or UV lasers. We analyzed the role of different key parameters, including size, aspect ratio and type of substrate. This study reveals that, unlike thermal or UV processing, visible light can induce the formation of sub-wavelength domains of crystalline silicon (c-Si) at the interior of the pillar amorphous (a-Si) matrix, which act as secondary optical antennas, providing unprecedented level of spatial and temporal control of lattice reconstruction in regions that cannot be accessed by conventional methods.

**Description of the system**

Amorphous silicon can undergo crystallization below its melting temperature, with the obvious advantage that the overall shape of the silicon structure is preserved. (16-18) Here we investigate the possibility to induce heat generation and control the spatial crystallization in amorphous silicon nanostructures by directly exploiting the coupling of light with optical resonances. Amorphous-Si nanostructures are characterized by low thermal conductivity (1.5 W/mK). Our reference model is represented by a silicon pillar (Si-P), vertically grown onto a given substrate and surrounded by air. The thermal effects induced by different excitation wavelengths (UV, 266 nm and Visible, 532 nm) are compared with those that are usually obtained by solid phase epitaxial regrowth (SPER), (19) by evaluating the evolution of the induced crystallization, $X_C$ (*see* Methods and Supplementary Material S1). The role of the aspect ratio is also studied over a wide range of values. The diameter of the Si-Ps ranged from 60 to 140 nm and the height varied from 150 to 640 nm. Light excitation is conveyed through the Si-P in form of a Gaussian beam, focused by a 0.9 NA optical objective. In typical simulations, the light fluence was set to 0.11 mW/$\mu$m$^2$, which is sufficient to induce local crystallization avoiding electronic ablation. The optically-driven crystallization kinetics was simulated by Finite Element Modelling (FEM) in a wide temporal range (from 1 fs to 25.3 hours, *see* Methods). We note that this interval is typically out of reach for methods based on molecular dynamics, which are limited to a few hundreds of nanoseconds. (20)

**Optical *vs* thermal annealing**

Figure 1 shows a comparison between the thermal effects that can be achieved by conventional thermal treatment (SPER) and those resulting from optical excitation. All the Si-Ps have the same size (height, h: 500 nm, diameter, d: 80 nm) and aspect ratio (6.25). In the simulation of conventional annealing the heat source temperature was set to 600°C at the bottom of SiP and heat flows from the substrate to the pillar. Thus, the crystallization process initiates at the basis of the pillar and propagates from the bottom to top (Figure 1a). On the contrary, UV irradiation induces the crystallization from top to bottom of the pillar (Figure 1b). This opposite behavior is due to the fact that the Visible light ($\lambda$=532 nm) efficiently couples to an optical resonance located in the middle of the pillar. On the other hand, UV light ($\lambda$=266 nm) dissipates its power more homogeneously over the pillar surface, thus the temperature is lower nearby the substrate and higher on the top of the pillar (Supplementary Material S2).

As a result, crystallization propagates isotropically in both directions (Figure 1c). These results indicate that optical excitation can be very useful to induce local crystallization in spatial regions of Si-Ps that are not accessible using conventional SPER. In general, for $\lambda$/d ratio values included between 5 and 10, the optical resonances are localized in the middle of the structure and can be directly excited by visible light, which could be interesting either in view of generating nanocrystalline regions embedded in an amorphous matrix, or for controlling the spatial distribution of dopants by creating high/low density regions.

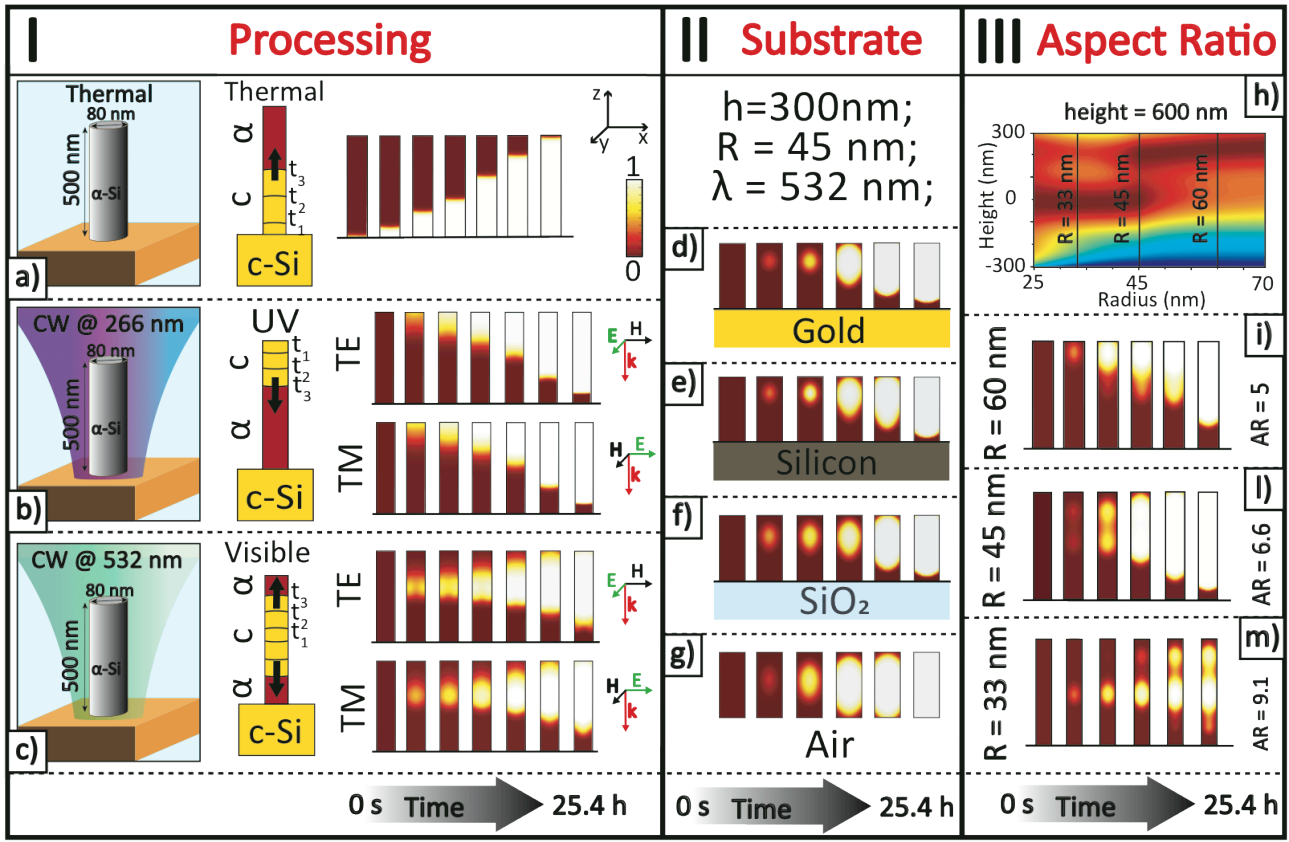

**Figure 1. Heat generation and propagation in Si nanopillars as a function of the type of processing (thermal annealing vs. photoirradiation), substrate and aspect ratio**. Panel I: The simulations of spatial and temporal evolution of the crystallization field, Xc, is simulated for a) solid phase epitaxial regrowth (annealing T=600°C), b) UV light laser irradiation (λ=266 nm) and c) Vis light laser irradiation (λ=532 nm) for a given aspect ratio (AR=6.25). In the case of the optical excitation the simulations for both TE and TM views are shown. Panel II: (d-g) Influence of the substrate on the spatial localization of Vis light laser induced crystallization (λ=532 nm) for a given aspect ratio (AR=3.3). Results for different AR values are reported in Supplementary Materials. Panel III: (h-m) Influence of AR. The temporal evolution of all the cases reported in the panels I-III was followed at different time intervals in a range between 0 and 91400 s. Details for each specific case is reported in Supplementary Materials S3).

**Role of the substrate**

Nanostructured antennas can be grown on different substrates. In order to clarify the role of the substrate in light-structure opto-thermal interactions and crystallization dynamics, we compare the use of different substrate materials. We chose gold, silicon and silica substrates as examples of conductors, semiconductors and insulators, respectively. The simulation of the behavior of free-standing SiPs was obtained by replacing the solid substrate with air. Figure 3 shows the results for a pillar of AR 3.3.

The resonance of the free standing pillar is located in the center of the structure, while the resonance of the pillar supported by $SiO_2$, Si and Au, is shifted toward the top. A similar trend is observed by decreasing the wavelength of the exciting radiation (*see* Supplementary Material S4).

On the other hand, when the AR of the pillar is larger than 5 the resonance is located far from the substrate, and the optical nature of the substrate does not significantly affect the optical resonance. No significant differences are observed in the crystallization dynamics. (*See* Supplementary Material S4)

**Role of aspect ratio**

Aspect ratio (AR) is a key parameter to determine the optical properties of a Si-P and, as a consequence, to control heat generation. Figure 1h shows the results of simulations of the opto-thermal response to a visible light irradiation ($\lambda$=532 nm) for Si-Ps with different aspect ratios (from 12 to 4.28), achieved by fixing the height of the Si-P to 600 nm and varying the radius from 25 to 70 nm. The color map represents the localization and intensity of the resonance-coupled heat source across the Si-P (hot spots). The crystallization kinetics of three different cases, corresponding to AR of 5, 6.6 and 9.1 was investigated and reported in Figures 2l-m, respectively. Figure 1i shows that the Si-Ps with the lowest AR (5) crystallize on the tip, which is a result analogous to that observed for UV-excited Si-Ps with higher (6.25) AR (see Figure 1b). By increasing the AR to 6.6 (Figure 1e) we observe the generation of two crystallization hot spots, which means that local temperature and crystallization efficiency are very similar in these regions. The hot spots are completely merged after 1.51 min and crystallization uniformly extends across the Si-P. For higher AR (9.1, Figure 1m) we observe the initial generation of a single crystalline hot-spot after 3.2 s. At 21 s an additional, weaker spot appears closer to the Si-P tip and both spots merge after 25.3 h. These results suggest that the local crystallization can be finely controlled by tuning the AR of a Si-P (*see* also Supplementary Material S5). The same irradiation source can produce hot spots located at different positions of the Si-P. Moreover, multiple hot spots can be generated at the same Si-P and maintained separated by stopping irradiation after a given time interval. The local temperature can vary from one hot spot to another, depending of the coupling efficiency between light and the optical resonances of the Si-P. Such a rich variety of effects and the precise spatial control of the degree of material crystallinity represent a powerful tool for management of the opto-thermal properties at the nanoscale.

**Light-induced crystallization dynamics**

Light-induced crystallization proceeds through a progressive modification of the degree of crystallinity of the material. The crystallization kinetics of Si-Ps was derived from the Johnson-Mehl-Avrami-Kolmogorov (JMAK) model modified by the Nakamura theory to account for non-isothermal conditions (*see* Methods). Crystallization follows a nucleation-growth process that is characterized by a stochastic nature. In this context, the terms $X_c$ (the degree of crystallization in the JMAK model) can be here regarded as the probability to find a portion of fully crystalline silicon in a given region of the Si-P. Panels a and b of Figure 2 show a comparison between the maxima value of $X_{c,max}$, which represents the presence/absence of crystalline grains, and the corresponding percentage of crystalline phases averaged over the whole Si-P volume ($X_{c,av}$), for a range of AR spanning from about 2.6 to 6.

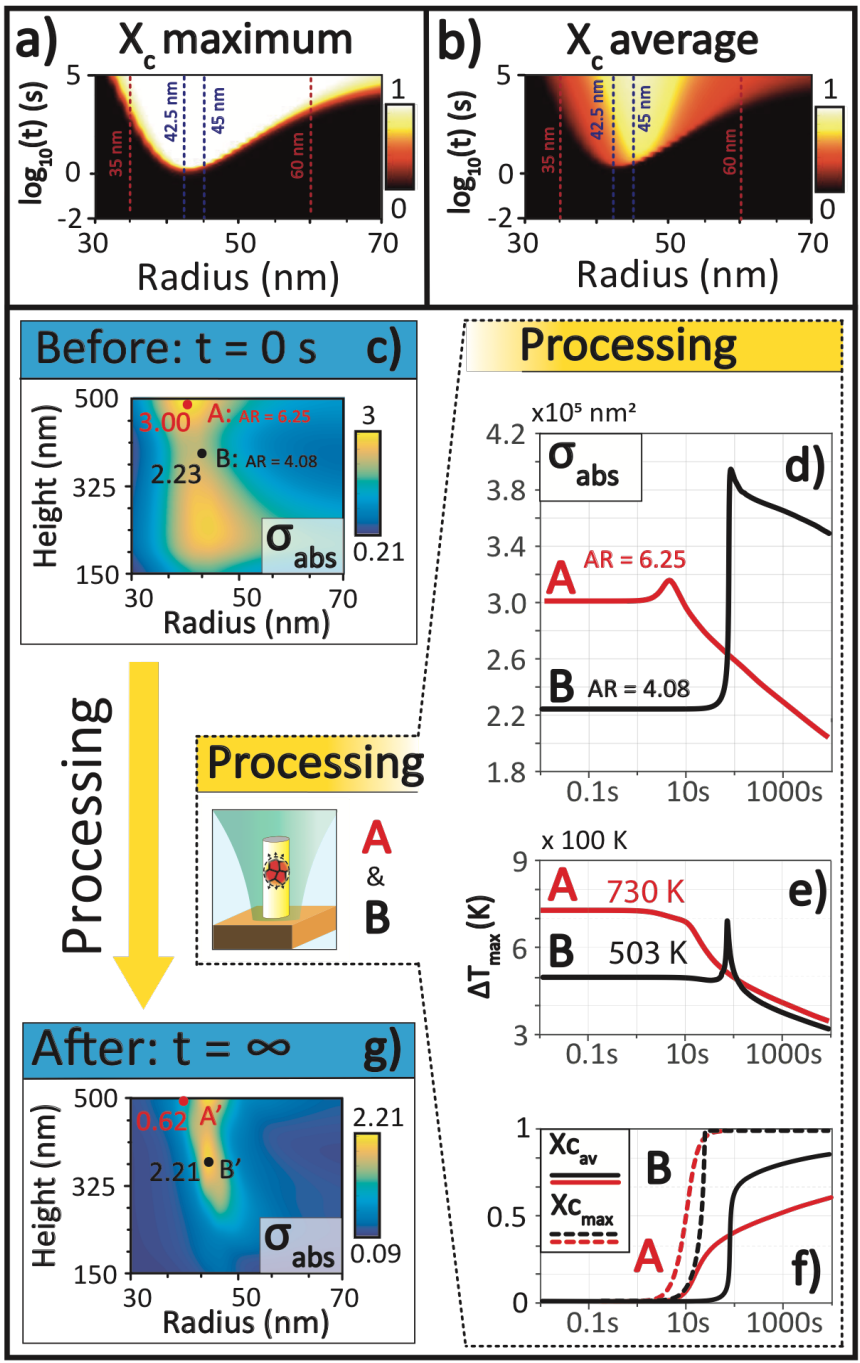

**Figure 2. Temporal evolution of optical, thermal and compositional properties of Si-Ps under visible light processing;** a) $X_{c,max}$ and b) $X_{c,ave}$ colour maps for different ARs and absolute absorption cross-section ($\lambda$=532 nm) color maps *before* (c) and *after* (g) irradiation for different ARs. Panels d, e and f show the time evolution of absolute absorption cross section, temperature enhancement and average Xc for two Si-Ps with different aspect ratios (AR sample A= 4.08, AR sample B=6.25). The positions of A and B samples on the absorption cross-section color maps before and after light processing are indicated by red and dark spots in panels c and g.

A direct comparison between the two maps reveals that $X_{c,ave}$ is always lower than $X_{c,max}$, which confirms that the light-driven crystallization is not homogeneous across the pillar and small differences in AR can strongly affect the crystallization efficiency and its spatial extension. In general, these data show a close correlation between structure-light coupling and the crystallization efficiency. The stronger is the pillar-light-coupling, the higher the temperature, and the faster the crystallization process. Whenever the diameter is either too small, or too large, the kinetics of crystallization is very slow, in the range of hundreds of seconds. The fastest crystallization rate, starting 1.5 s after laser irradiation, is observed for Si-Ps 42.5 nm in radius (AR: ~4.3). When the radius is slightly increased to 45 nm (AR:~ 4) the crystallization onset is similar (2 s), but $X_{c,ave}$ is significantly different (0.93 at t=25.3 h, while it is only 0.83 for the sample with the 4.3 AR after the same irradiation time). This suggests that the changes in the optical proprieties of nanopillars resulting from the growth of polycrystalline Si domains can deeply affect light-structure coupling and heat propagation, thus controlling the overall kinetics of the process.

This hypothesis was investigated by monitoring the evolution of the optical properties of Si-Ps with different aspect ratios as a function of laser processing time (λ=532 nm). Figure 2c shows a color map of the absorption cross-section of the amorphous Si-Ps before laser irradiation.

The highest absorption values are observed for Si-Ps with radii within 37.5-42.5 nm and lengths ranging from 450 to 500 nm, corresponding to aspect ratios varying from about 5.3 to 6.6. We will focus our attention on two Si-Ps with two different aspect ratios, 6.25 and 4, labeled as points A and B, respectively (from now on, samples A and B). Sample A corresponds to the highest absorption value $(3 \cdot 10^5$ nm$^2)$, whereas sample B is at the border of high/low light absorption region (absorption~$2.2 \cdot 10^5$ nm$^2$). Figures 2d-f monitor the evolution of absorption cross-section, temperature variation and $X_c$ as a function of irradiation time for both samples. As a result of amorphous-to-polycrystalline Si conversion, at the end of the irradiation process A and B are transformed into the corresponding A' and B' samples, characterized by different optical properties (Figure 2g). In particular, the optical absorption of sample with AR= 6.25 moves from in-resonance (A) to out-of-resonance (A') conditions. The absorption cross section of this sample passes through a maximum after 4.5 s ($X_c$ =0.73) and decreases to lower values ($\sim 0.6 \cdot 10^5$ nm$^2$) for prolonged irradiation. On the other hand, sample with AR=4 moves from low absorption values (B) and reaches its maximum ($\sim 4 \cdot 10^5$ nm$^2$) after 86.7 s (B'), passing through a steep increase at$\sim$ 80 s. The evolution of temperature (Figure 2e) shows that both systems reach the thermal equilibrium after few hundreds of nanoseconds from the beginning of irradiation. The transient equilibrium temperature reaches 730 and 503 K for system A and B, respectively. Interestingly, we note that after 86.7 s, in precise correspondence with the narrow peak observed in optical absorption (Figure 2d), sample B exhibits a very sharp peak, as the temperature reaches 700 K and rapidly decays. At this stage $X_{c,max}$ is 0.99, but $X_{c,ave}$ is only 0.35, which confirms that only few crystalline grains, embedded in an amorphous matrix, can generate resonances that can rapidly boost the local temperature. At the end of laser processing (25.3 h), the $X_{c,ave}$ value is 0.76, while $X_{c,max}$ is 1 (Figure 2f). The B pillar exhibits a lower conversion rate in the amorphous phase. However, when crystallization begins, light-structure coupling increases abruptly and the kinetics becomes much faster in comparison to the A-counterpart. Overall, the crystallization kinetics of B is faster than A. The key role of the optical resonances is further discussed in Supplementary Material S6, which shows that non-resonant UV processing is not able to produce any secondary antenna-effects.

The absorbed power $Q_e$ as a function of time for the two systems, A and B is shown in Figure 3. By multiplying the absorbed power density by the Xc field and integrating over the pillar volume, it is possible to decouple the power absorbed by the crystalline phase from that of the amorphous phase. At t=4.5 s, as a result of the appearance of the first crystalline seeds in the pillar with AR ~4 (sample A), there is a strong enhancement in the absorbed power of the crystalline inclusion (blue curve). A similar trend can be observed for the pillar with AR ~6 but at t=80 s. In general, two evolution regimes, can be distinguished:

I) a fast regime, where the system is quickly kept out of equilibrium in a metastable state. This persists for few seconds after crystallization of the first seeds (light blue region in Figure 3a and b). In particular, it lasts for 4 s (from 0.4 s to 4.4 s) in A and for 62 s (from 31 s to 93 s) in B.

II) a slow regime where the system moves from the metastable towards the equilibrium state. This persists until the whole pillar has been crystallized (light orange region in Figure 3a and b).

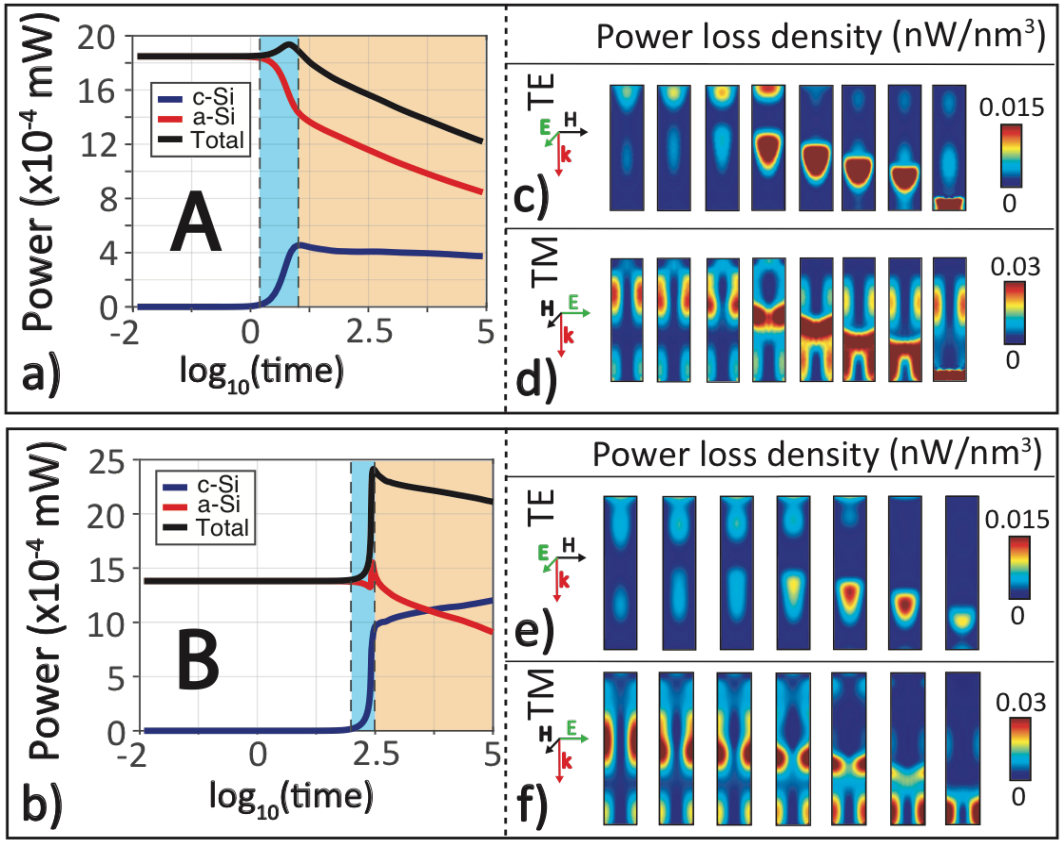

**Figure 3. Time evolution of the absorbed power and spatial evolution of power loss density:** total absorbed power (black line), power absorbed by the portion of a-Si (red line) and by the portion of c-Si (blue line) for samples A (panel A) and (panel B). The light blue region shows the time interval of the fast regime, the light orange region shows the time interval of the slow regime. The spatial evolution of the power loss density for different times is reported in panel c) and d).

Figures 3c-f show the temporal and spatial evolution of power dissipation for A and B at two different polarization of the electric field (TE and TM). The crystallization onset results in a strong localization of the absorbed photons, which takes place in the amorphous region nearby the inclusion. Different spatial distribution can be observed from the TE and TM sections. As the crystalline phase increases, the enhanced absorption remains localized in the a-Si under the propagating front. Such an efficient localization reinforces the hypothesis that the crystalline inclusions may act as secondary opto-thermal antenna within the pillar.

Figure 4 shows a summary of the light-induced crystallization dynamics in Si-NPs. The light-induced local crystallization determines a change of the optical properties in the irradiated area, which in turns promotes further light absorption. This secondary antenna, made of nanocrystalline silicon embedded within an amorphous silicon matrix, boosts crystallization until the focal region is displaced out of the resonance condition. At this stage, the optical absorption is reduced and the heat generation process undergoes self-limitation. The dynamic modification of the opto-thermal properties of the system as a function of irradiation time offers a real-time feedback mechanism for heat propagation, which allows a precise temporal control of crystallization and lattice reconstruction processes. We note that the change of refractive index of silicon as a function of the temperature reached upon light irradiation is negligible for the amorphous phase, whereas becomes significant in the case of nanocrystals (21). Results of simulations carried out by including the variation of the refractive index of the material as a function of temperature show qualitatively similar results of those obtained by neglecting this effect, although the latter approximation yields underestimation of light absorption peak and $X_{c,ave}$ (*see* Supplementary Material S7).

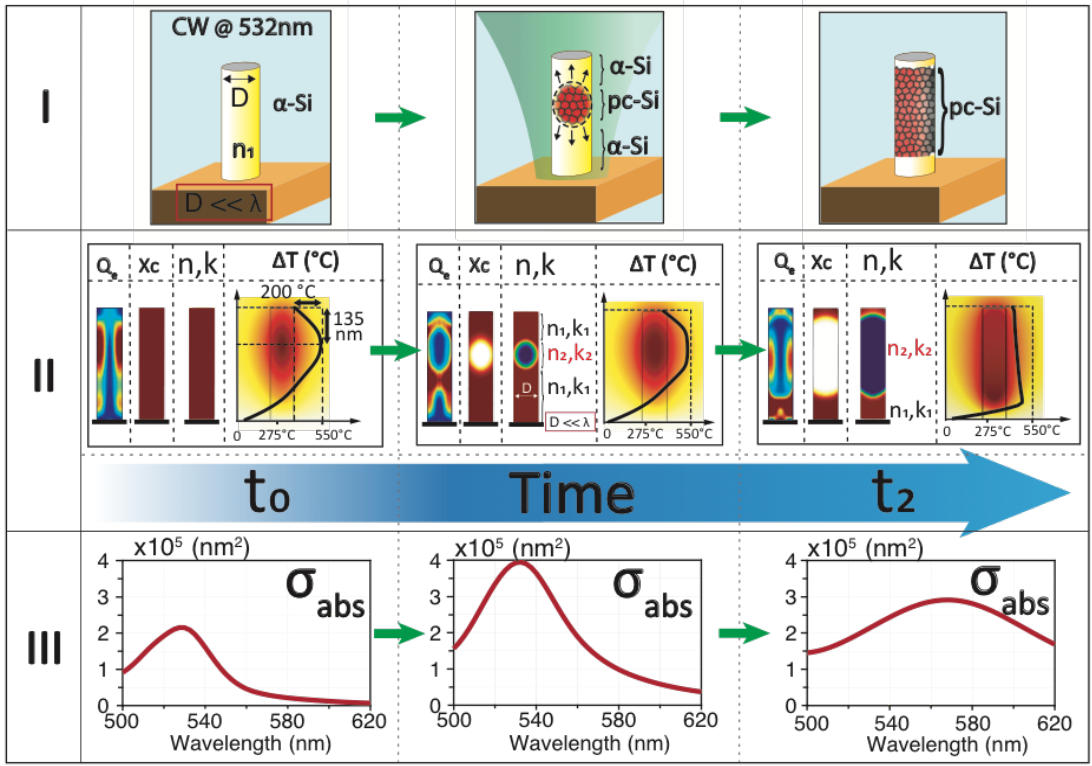

**Figure 4. Dynamics of light-induced crystallization process.** Panel I shows a schematic representation of the time evolution for crystallization of a resonant silicon pillar. Panel II shows the simulations of the absorbed power, $X_c$, complex refractive index, and temperature at different time, as the process takes place. Panel III shows the modification of the absorption cross-section, at different stages of the crystallization.

**Conclusions**

This study demonstrates that a rational exploitation of the optical resonances in silicon nanostructures allows the precise generation of localized heating, which can be controlled by size, substrate and aspect ratio. In particular, visible lasers can be utilized to drive local crystallization and lattice reconstruction in regions that cannot be reached by other methods, offering a powerful tool for developing new photonic devices based on silicon nanostructures (25-27). The theoretical model developed in this work is able to predict the spatial evolution of the light-driven opto-thermal effects over an extended temporal range, which spans from fs to hours. The model also revealed that the crystallization dynamics can give rise to secondary antennas embedded into the primary nanostructures, which broadens the horizons of light harvesting and thermal management at the nanoscale. This approach can be extended to many different materials and thermally-activated processes, including laser resonant printing (28), diffusion-based doping of semiconductors, nanofabrication of all-dielectric and hybrid devices for enhanced vibrational spectroscopy (29-31) and photothermal therapy.

**Methods**

The whole process of light absorption, heat transfer and phase change was investigated by means of numerical simulation based on Finite Element Analysis implemented in Comsol Multiphysics (Comsol Inc., Burlington, MA). The change of optical proprieties in the pillars is caused by the silicon crystallization. Amorphous silicon can undergo solid crystallization by heating below its melting point. To describe the non-isothermal crystallization kinetics, a development of the Johnson-Mehl-Avrami-Kolmogorov (JMAK) model has been used. The JMAK model represents a simple and widespread way to describe the phase change kinetics in an isothermal process. For a given material and a fixed temperature T, the percentage of crystallized substance $X_c$ is described by the following equation:

$$X_c = 1 - exp(-k \cdot t^n) \qquad (1)$$

where $k$ is called isothermal constant and $n$ the Avrami number. If temperature is not constant, eq. 1 is no longer correct. Nakamura *et al*. (22) provided a more general theory, extending the Avrami treatment for non-isothermal processes. For a given temperature $T(\boldsymbol{r}, t)$, the $X_c(\boldsymbol{r}, t)$ can be obtained by solving the following differential equation:

$$\frac{\partial X_c(\boldsymbol{r}, t)}{\partial t} = n \cdot K(T(\boldsymbol{r}, t)) \cdot (1 - X_c(\boldsymbol{r}, t)) \left[ ln \left( \frac{1}{1 - X_c(\boldsymbol{r}, t)} \right) \right]^{\frac{n-1}{n}} \qquad (2)$$

where $n$ is the exponent of the JAM equation, and $K(T(\boldsymbol{r}, t))$ is called non-isothermal constant, which can be related to the JAM crystallization constant $k(T)$ by the following expression:

$$K(T) = k(T)^{1/n} \quad (3)$$

For the spatial description of crystallization, we assume that if a temperature gradient is present in the material, the Nakamura model can still be used, but now $X_c$ is a space-dependent quantity, assuming any possible values between 0 and 1. Given the stochastic nature of nucleation-growth process (23), we attribute to $X_c$ a statistical interpretation, as the probability to find a crystalline region in a given point of the system.

The temperature field $T(\boldsymbol{r}, t)$ is iteratively recalculated every time step, solving the time-dependent heat transfer equation:

$$\rho(T, X_c) \cdot C_p(T, X_c) \cdot \frac{\partial T}{\partial t} - \nabla \cdot (k(T, X_c) \cdot \nabla T) = Q_e \qquad (4)$$

where $\rho(T, X_c)$ is the density, $C_p(T, X_c)$ is the heat capacity at constant pressure, $k(T, X_c)$ is the thermal conductivity and $Q_e$ is the heat source. All the coefficients involved in eq. 4 vary with temperature and with $X_c(\boldsymbol{r}, t)$. See supporting information for further details.

The heat source $Q_e$ is obtained from the EM absorption, described by the expression:

$$Q_e = \frac{1}{2} \varepsilon_0 Re \left\{ \frac{\omega \varepsilon''(\omega, X_c)}{4\pi} \boldsymbol{E} \cdot \boldsymbol{E}^* \right\} \quad (5)$$

where $\boldsymbol{E}$ is the complex electric field, and $\varepsilon''(\omega, X_c)$ is the imaginary part of the relative dielectric permeability. The electromagnetic absorption has been considered as the only heat source in the system. The numerical solution of Maxwell's equations provides at each time step the values for electric and magnetic fields $\boldsymbol{E}, \boldsymbol{H}$ in all the points of the system. While the system undergoes crystallization, the electric and magnetic fields are recalculated and the losses are updated. This enables to follow the dynamical modification of optical proprieties, while crystallization takes place.

The $X_c$ dependence of optical and thermal constants are detailed in the Supplementary Material S1.

The modification of optical properties as a function of granularity is negligible in comparison to the variation of optical proprieties from the amorphous- to-crystalline phase (23). For simplicity, in the model we consider this evolution is accompanied by the corresponding modification of the optical properties of the Si-P, which in turn affects light absorption and heat generation. As a result, the overall kinetics of heat propagation and its spatial extension depend on this mutual feedback. Thus, in evaluating the dynamics of crystallization and heat propagation, we must consider that the optical properties of the pillars are changing as a function of the crystallization degree that has been reached at a given time and the temperature distribution is not constant across the pillar (*see* Supplementary Material S1).

**References**

(1) Emma, P. G., & Kursun, E. Is 3D chip technology the next growth engine for performance improvement? *IBM J. Res. Dev.* **52**, 541–552 (2008).

(2) Cui, Y. & Lieber, C. M. Functional nanoscale electronic devices assembled using silicon nanowire building blocks. *Science* **291**, 851–853 (2001).

(3) Wong, H.S.P., & Salahuddin, S. Memory leads the way to better computing. *Nature Nanotechnology* **10**, 191–194 (2015).

(4) Yoo, S. J. B., Guan, B., & Scott, R.P. Heterogeneous 2D/3D photonic integrated microsystems. *Nature Microsystems & Nanoengineering* **2**, 16030 (2016).

(5) Keren-Zur, S., & Ellenbogen, T. A new dimension for nonlinear photonic crystals. *Nature Photonics* **12**, 575–577 (2018).

(6) Leung, S., Zhang, Q., Xiu, F., Yu, D., Ho, J. C., *et al*. Light Management with Nanostructures for Optoelectronic Devices. *Journal of Physical Chemistry Letters* **5**, 1479–1495 (2014).

(7) Mirabella, S., De Salvador, D., Napolitani, E., Bruno, E. & Priolo, F. Mechanisms of boron diffusion in silicon and germanium. *Journal of Applied Physics* **113**, 031101 (2013).

(8) Williams, J.S. Solid phase epitaxial regrowth phenomena in silicon. *Nuclear Instruments and Methods in Physics Research* **209**, 219-228 (1983).

(9) Misra, N., Xu, L., Pan, Y., Cheung, N., Grigoropoulos, C. P. Excimer laser annealing of silicon nanowires. *Applied Physics Letters* **90**, 111111 (2007).

(10) Ucjikoga, S. Low-Temperature Polycrystalline Silicon Thin-Film Transistor Technologies for System-on-Glass Displays. *MRS Bulletin* **27**, 881 (2002).

(11) Caldarola, M., Albella, P., Cortés, E., Rahmani, M., Roschuk, T., et al. Non-plasmonic nanoantennas for surface enhanced spectroscopies with ultra-low heat conversion. *Nature Communications* **6**, 7915 (2015).

(12) Olson, G.L., Roth, J.A. Kinetics of solid phase crystallization in amorphous silicon. *Materials Science Reports* **3**, 1-77 (1988).

(13) Danesi, S., Gandolfi, M., Carletti, L., Bontempi, N., De Angelis, *et al*. Photo-induced heat generation in non-plasmonic nanoantennas. *Physical Chemistry Chemical Physics* **20**, 15307-15315 (2018).

(14) Bontempi N., Vassalini I., Danesi S., Ferroni M., Donarelli M. *et al*. Non-Plasmonic SERS with Silicon: Is It Really Safe? New Insights into the Optothermal Properties of Core/Shell Microbeads. *Journal of Physical Chemistry Letters* **9**, 2127-2132 (2018)

(15) Zograf, G. P., Petrov, M. I., Zuev, D. A., Dmitriev, P. A., Milichko, V. A. Resonant Nonplasmonic Nanoparticles for Efficient Temperature-Feedback Optical Heating. *Nano Letters* **17**, 2945–2952 (2017).

(16) Hong, W. E., and Ro, J.S. Kinetics of solid phase crystallization of amorphous silicon analyzed by Raman spectroscopy. *Journal of Applied Physics* **114**, 073511 (2013).

(17) Wang, L., Rho, Y., Shou, W., Hong, S., Kato, K., et al. Programming nanoparticles in multiscale: optically modulated assembly and phase switching of silicon nanoparticle array. *ACS Nano*. **12**, 2231-2241 (2018).

(18) Makarov, S. V., Zalogina, A. S., Tajik, M., Zuev, D. A., Rybin, M. V., et al. Light-Induced Tuning and Reconfiguration of Nanophotonic Structures. *Laser Photonics Review* **11**, 1700108 (2017).

(19) Morarka, S., Rudawski, N. G., Law, M. E., Jones, K. S., and Elliman, R. G. Modeling two-dimensional solid-phase epitaxial regrowth using level set methods. *Journal of Applied Physics* **105**, 053701 (2009).

(20) In, J. B., Xiang, B., Hwang, D. J., Ryu, S. G., Kim, E., et al. Generation of single-crystalline domain in nano-scale silicon pillars by near-field short pulsed laser. *Applied Physics A* **114**, 277-285 (2014).

(21) Jellison Jr, G. E., and Burke, H. H. The temperature dependence of the refractive index of silicon at elevated temperatures at several laser wavelengths. *Journal of Applied Physics* **60**, 841 (1986).

(22) Nakamura, K., Watanabe, T., Katayama, K. & Amano, T. Some aspects of nonisothermal crystallization of polymers. Relationship between crystallization temperature, crystallinity, and cooling conditions. *Journal of Applied Polymer Science* **16**, 1077-1091 (1972).

(23) Spinella C., Lombardo S. and Priolo F. Crystal grain nucleation in amorphous silicon. *Journal of Applied Physics* **84**, 5383 (1998).

(24) Palik, E. D. *Handbook of Optical Constants of Solids*. *Vol. 1*, (Academic Press, 2012).

(25) Priolo, F., Gregorkiewicz, T., Galli M. & Krauss, T. F. Silicon nanostructures for photonics and photovoltaics. *Nature Nanotech*. **9**, 19-32 (2014).

(26) Vella A., Shinde D., Houard J., Silaeva E., Arnoldi L. *et al*. Optothermal response of a single silicon nanotip. *Phys. Rev. B* **97**, 075409 (2018).

(27) Van Laer R., Kuyken B, Van Thourhout D & Baets R. Interaction between light and highly confined hypersound in a silicon photonic nanowire. *Nature Photon*. **9**, 199-203 (2015).

(28) Zhu X., Yan W., Levy U., Asger Mortensen N., Kristensen A. Resonant laser printing of structural colors on high-index dielectric metasurfaces. *Sci*. *Adv*. **3**, e1602487 (2017).

(29) Alessandri I & Lombardi, J. R. Enhanced Raman scattering with dielectrics. *Chem*. *Rev*. **116,** 14921-14981 (2016).

(30) Dmitriev P. A., Baranov D. G. Milichko V. A., Makarov S. V., Mukhin I. S. *et al*. Resonant Raman scattering from silicon nanoparticles enhanced by magnetic response. *Nanoscale* **8**, 9721-9726 (2016).

(31) Frizyuk K., Hasan M., Krasnok A., Alú A. & Petrov M. Enhancement of Raman scattering in dielectric nanostructures with electric and magnetic Mie resonances, *Phys*. *Rev*. *B* 97 085414 (2018).

**Acknowledgments**

We thank Prof. C. De Angelis, University of Brescia, for a critical review of the manuscript and valuable discussions. This work was carried out in the framework of the project: "Microsfere adattative per il monitoraggio e l'abbattimento di inquinanti persistenti-MI ADATTI E L'ABBATTI" supported by INSTM and Regione Lombardia.

**Corresponding Author**

*Ivano Alessandri. E-mail: ivano.alessandri@unibs.it

**Author Contributions**

S. D. and I. A conceived the study. S. D. performed the numerical simulation of the opto-thermal properties. All the authors discussed the results and contributed to the final version of the manuscript.

**Competing interests**

The authors declare no competing financial interest.